\documentclass[iop]{emulateapj}

\begin{document}

\title{IONIZATION SOURCE OF A MINOR-AXIS CLOUD IN THE OUTER HALO OF M82}

\author{K. Matsubayashi\altaffilmark{1},
H. Sugai\altaffilmark{2},
A. Shimono\altaffilmark{2},
T. Hattori\altaffilmark{3},
S. Ozaki\altaffilmark{4},
T. Yoshikawa\altaffilmark{5},
Y. Taniguchi\altaffilmark{1},
T. Nagao\altaffilmark{5, 6},
M. Kajisawa\altaffilmark{1, 7},
Y. Shioya\altaffilmark{1}, and
J. Bland-Hawthorn\altaffilmark{8}
}
\altaffiltext{1}{Research Center for Space and Cosmic Evolution, Ehime
University, Bunkyo-cho 2-5, Matsuyama, Ehime 790-8577, Japan: 
kazuya@cosmos.phys.sci.ehime-u.ac.jp}
\altaffiltext{2}{Institute for the Physics and Mathematics of the Universe,
The University of Tokyo, Kashiwanoha 5-1-5, Kashiwa, Chiba 277-8583, Japan}
\altaffiltext{3}{Subaru Telescope, National Astronomical Observatory of Japan,
650 North A'ohoku Place, HI, 96720, USA}
\altaffiltext{4}{National Astronomical Observatory of Japan, 2-21-1, Osawa,
Mitaka, Tokyo 181-8588, Japan}
\altaffiltext{5}{Department of Astronomy, Kyoto University, Kitashirakawa
Oiwakecho, Sakyo-ku, Kyoto 606-8502, Japan}
\altaffiltext{6}{The Hakubi Center for Advanced Research, Kyoto University,
Yoshida-Ushinomiya-cho, Sakyo-ku, Kyoto 606-8302, Japan}
\altaffiltext{7}{Graduate School of Science and Engineering, Ehime University,
Bunkyo-cho 2-5, Matsuyama, Ehime 790-8577, Japan}
\altaffiltext{8}{Sydney Institute for Astronomy, School of Physics, University
of Sydney, Camperdown, NSW 2006, Australia}

\begin{abstract}

The M82 `cap' is a gas cloud at a projected radius of 11.6 kpc along the minor 
axis of this well known superwind source.
The cap has been detected in optical line emission and X-ray emission and
therefore provides an important probe of the wind energetics.
In order to investigate the ionization source of the cap, we observed it with
the Kyoto3DII Fabry-Perot instrument mounted on the Subaru Telescope.
Deep continuum, H$\alpha$, [\ion{N}{2}]$\lambda$6583/H$\alpha$, and
[\ion{S}{2}]$\lambda\lambda$6716,6731/H$\alpha$ maps were obtained with
sub-arcsecond resolution.
The superior spatial resolution compared to earlier studies reveals a number of
bright H$\alpha$ emitting clouds within the cap.
The emission line widths ($\lesssim 100$ km s$^{-1}$ FWHM) and line ratios in 
the newly identified knots are most reasonably explained by slow to moderate
shocks velocities ($v_{\rm shock}$ = 40--80 km s$^{-1}$) driven by a fast wind
into dense clouds.
The momentum input from the M82 nuclear starburst region is enough to produce
the observed shock.
Consequently, earlier claims of photoionization by the central starburst are
ruled out because they cannot explain the observed fluxes of the densest knots
unless the UV escape fraction is very high ($f_{\rm esc} >$ 60\%), i.e., an
order of magnitude higher than observed in dwarf galaxies to date.
Using these results, we discuss the evolutionary history of the M82 superwind.
Future UV/X-ray surveys are expected to confirm that the temperature of the gas
is consistent with our moderate shock model.

\end{abstract}


\section{INTRODUCTION}

Superwinds are galaxy scale outflows, caused by supernovae in nuclear starburst
regions or active galactic nuclei (AGNs).
They are so powerful that interstellar matter within the galaxies is blown out. 
Some of the material may escape to the intergalactic or group medium, while
some of the material may be recycled throughout the galactic halo
\citep{Cooper:2008}.
Superwinds are expected to quench star-formation activity (feedback) and to
enrich the external medium with new metals.
Generally, galactic winds are diffuse and difficult to observe.
M82, one of the nearest starburst galaxies \citep[3.63 Mpc, ][]{Freedman:1994},
is one of the most well known examples of the superwind phenomenon. 
Its large inclination angle \citep[$i \sim$ 80$^{\circ}$: ][]
{Lynds:1963, McKeith:1995} and proximity allow us to see many details of the
wind phenomenon far from the galactic plane.
The source has been observed in hot gas ($T \sim 10^6$ K; e.g.,
\citealt{Cappi:1999}), ionized gas ($T \sim 10^4$ K; e.g.,
\citealt{McCarthy:1987, Bland:1988}), and molecular gas ($T \sim 10^2$ K;
e.g., \citealt{Nakai:1987, Walter:2002}). 

The kinematics and ionization of the wind material over the inner few
kiloparsecs have been investigated in detail.
\citet{McKeith:1995} and \citet{Shopbell:1998} modeled the outflow structure
using position-velocity diagrams in optical emission lines.
The emission line ratios of the inner region indicate that photoionization by
the nuclear starburst plays a significant role in the excitation
\citep{McCarthy:1987,Shopbell:1998}.
In recent years, new observational methods such as integral field spectroscopy
(e.g., \citealt{Westmoquette:2009a, Westmoquette:2009b}) and spectropolarimetry
(e.g., \citealt{Yoshida:2011}) have revealed its more intricate structure.

Our goal is to shed light on processes behind large-scale galactic winds.
Very little is known about their total extent, energetics and importance in the
context of galaxy evolution.
By studying the most spatially extended emission, we can obtain a better
understanding of the total kinetic energy of the wind.
There are many questions that remain unanswered for M82's outflow. 
How old is the wind and how far does it extend?
Is it powered by radiation pressure or wind pressure, or a combination of both?
Is the source of energy impulsive or sustained over many dynamical times?
Is most of the outflowing material swept up or entrained from the disk?
Does the wind material escape the galaxy or fall back to the disk?
To have any chance of answering these questions, we need a better understanding
of the most basic properties of the large-scale wind.

The most distant gas cloud in M82 is the `cap' originally discovered in
H$\alpha$ and X-ray emission at a radius of 11.6 kpc along the minor axis of
M82 \citep{Devine:1999, Lehnert:1999}.
Strong UV emission provides evidence for reflecting dust in the cloudlets that
make up the cap \citep{Hoopes:2005}.
The metal abundances of O, Ne, Mg, Si, and Fe of X-ray emitting gas in the cap
suggest that most of the metals arise from a circumnuclear starburst dominated
by Type II supernovae \citep{Tsuru:2007}.

We now show that the dominant ionization source in the cap provides an
important clue to the wind's origin and history.
\citet{Lehnert:1999} suggested the cap is either photoionized by UV photons
from the nuclear starburst region or by a shock being driven by the hot wind
into a dense halo cloud, or a combination of both.
The X-ray observations already support the idea that the wind reaches the
distance of the cap, but are the optical emission line diagnostics consistent
with a wind-driven shock?

Therefore, in order to obtain emission line intensity map and line ratio maps
at high spatial resolution, we carried out Fabry-Perot observations of M82's
cap with the Subaru Telescope.
This combination enables us to detect weak emission with a larger field of view
than that of integral field spectroscopy.
Through a comparison of the observed line ratios and those calculated by
photoionization and shock models, we discuss the ionization source of the M82
cap and a likely evolution history for the large-scale galactic wind.

\section{OBSERVATIONS AND DATA REDUCTION}

\begin{figure}
\epsscale{.90}
\plotone{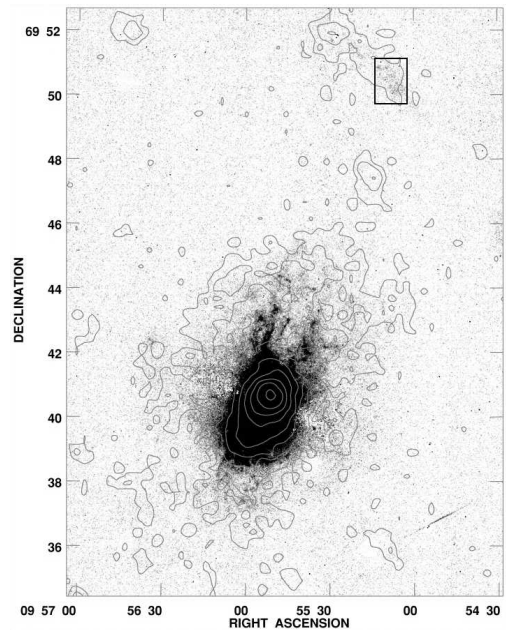}
\caption{H$\alpha$ (greyscale) and X-ray (contours) images of the whole M82
\citep{Lehnert:1999}.
The field of view of Figure \ref{fig:image-3dii} panels is displayed with a
rectangle.
\label{fig:image-m82-whole}}
\end{figure}

On 2011 November 22, we observed the central part of the M82 cap,
$\sim$10\arcmin\, N of the nucleus of M82, with the Kyoto3DII Fabry-Perot mode
\citep{Sugai:2010} mounted on the Cassegrain focus of the Subaru Telescope.
Figure \ref{fig:image-m82-whole} displays the position of the cap relative to
the M82 center, and indicates the region where we observed in this observation.
This mode uses an ET-50 etalon manufactured by Queensgate Instruments.
The field of view is $\sim$1\arcmin.9 and the pixel scale is 0\arcsec.112
pixel$^{-1}$ after 2 $\times$ 2 on-chip binning.
The spectral resolution $R \approx 348$ corresponds to 19 \AA\, at 6598.95\AA.
We obtained 14 object frames for H$\alpha$ +
[\ion{N}{2}]$\lambda\lambda$6548,6583, five for
[\ion{S}{2}]$\lambda\lambda$6716,6731, and two for the off bands.

The observed wavelengths at the field centres are summarized in Table
\ref{tb:obs-wavelength}.
The exposure time for each frame was 300 seconds.
We also observed a standard star EGGR247 for flux calibration 
\citep{Bland-Hawthorn:1995}.
Bias subtraction and flat fielding were performed for the target and standard
star frames.
Because the center wavelength in Fabry-Perot observations depends on the
distance from the center of the field of view, simple sky subtraction results
in some residuals due to sky emission lines.
We measured sky emission fluxes in blank regions of the object frames, and
subtracted it from the regions at the same distance from the center.
Flux calibration and distortion correction were carried out for the target
frames.
We used a spectrum catalog of \citet{Oke:1990} for flux calibration for each
wavelength setting.
The positional offsets among the object frames were detected, because the
Cassegrain auto guider was unavailable due to repairs and we did not use it in
this observation run.
We corrected the offsets by using the stars in the target frames.
We matched the spatial resolution of the target frames to the worst one,
0\arcsec.9, and carried out 4 $\times$ 4 binning, resulting in the pixel scale
of 0\arcsec.45 pixel$^{-1}$.

\begin{deluxetable}{cc}
\tablecaption{Observed wavelengths for the M82 cap
\label{tb:obs-wavelength}}
\tablewidth{0pt}
\tablehead{
\colhead{band} & \colhead{wavelength [\AA]}
}
\startdata
H$\alpha$ + [\ion{N}{2}]$\lambda\lambda$6548,6583 & 6546,
6554\tablenotemark{a}, 6562\tablenotemark{a}, 6570\tablenotemark{a},
6578\tablenotemark{a}, 6586\tablenotemark{a}, 6594\tablenotemark{a}, 6602 \\
{[\ion{S}{2}]}$\lambda\lambda$6716,6731 & 6714, 6722, 6730, 6738, 6746 \\
continuum & 6656\tablenotemark{a} 
\enddata
\tablenotetext{a}{Two exposures were performed at these wavelengths.}
\tablecomments{These are the observed wavelengths at the center of our field of
view.}
\end{deluxetable}

Because of the relatively low spectral resolution, H$\alpha$ and
[\ion{N}{2}]$\lambda\lambda$6548,6583 were blended in H$\alpha$+[\ion{N}{2}]
band.
We fitted these lines pixel by pixel with the transmission curve of the
Fabry-Perot interferometer \citep[Airy function: ][]{Bland:1989} and decomposed
them (see \citealt{Matsubayashi:2009}).
Better wavelength sampling than that of the previous observation enables us to
find the best velocity center.
We fitted the emission line fluxes at each velocity, from 100 km s$^{-1}$ to
700 km s$^{-1}$, and selected the velocity for which the fitting residual is
the smallest.
For line decomposition, we assumed that the [\ion{N}{2}]$\lambda$6548 flux is
one-third of [\ion{N}{2}]$\lambda$6583 flux, and that the velocity centers of
H$\alpha$ and [\ion{N}{2}]$\lambda\lambda$6548,6583 are same.

The velocity dispersion is fixed to 0 km s$^{-1}$ because it is much smaller
than the spectral resolution of the instrument.
This assumption is reasonable, since the observed velocity dispersion of
H$\alpha$ at the cap is small ($\sim$100 km s$^{-1}$; \citealt{Devine:1999}).
The same fitting was performed for the [\ion{S}{2}] band data.
However, the [\ion{S}{2}]$\lambda$6716/[\ion{S}{2}]$\lambda$6731 cannot be
determined well, because the wavelength difference between these two lines is
smaller than the spectral resolution in this observation.
Therefore we only use the total flux of [\ion{S}{2}]$\lambda\lambda$6716,6731
in this study.
Our 1$\sigma$ detection limit in H$\alpha$ surface brightness is estimated to
be 6.5 $\times$ 10$^{-18}$ erg cm$^{-2}$ s$^{-1}$ arcsec$^{-2}$ or an emission
measure of roughly 1 Rayleigh (3.3 cm$^{-6}$ pc) at a temperature of 10$^4$ K.
We adopt a distance of 3.63 Mpc to M82 \citep{Freedman:1994}.

\section{RESULTS}
\label{sec:results}

\begin{figure}
\epsscale{1.20}
\plotone{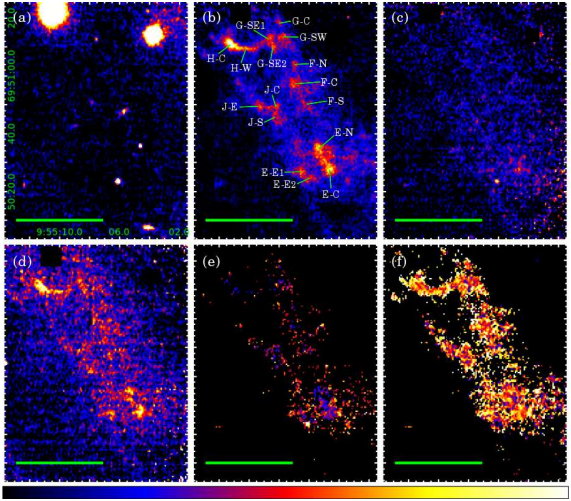}
\caption{(a) Continuum surface brightness, (b) H$\alpha$ intensity, (c)
[\ion{N}{2}]$\lambda$6583 intensity, (d) [\ion{S}{2}]$\lambda\lambda$6716,6731
intensity, (e) [\ion{N}{2}]/H$\alpha$ ratio, and (f) [\ion{S}{2}]/H$\alpha$ 
ratio maps of a part of the M82 cap.
The color scales are 0--3.3 $\times$ 10$^{-19}$ erg cm$^{-2}$ s$^{-1}$
\AA$^{-1}$ (0.45 arcsec)$^{-2}$,
0--2.5 $\times$ 10$^{-17}$ erg cm$^{-2}$ s$^{-1}$ (0.45 arcsec)$^{-2}$,
0--1.5 $\times$ 10$^{-17}$ erg cm$^{-2}$ s$^{-1}$ (0.45 arcsec)$^{-2}$,
0--1.5 $\times$ 10$^{-17}$ erg cm$^{-2}$ s$^{-1}$ (0.45 arcsec)$^{-2}$,
0--1.0, and 0--1.0 for the continuum, H$\alpha$, [\ion{N}{2}],
[\ion{S}{2}], [\ion{N}{2}]/H$\alpha$, and [\ion{S}{2}]/H$\alpha$ maps,
respectively.
Two bright stars appeared at upper side in the panel (a) are masked in the
panels (b)--(f).
North is up and east is left.
Bar in lower left in each panel represents 30\arcsec, which corresponds to 0.5
kpc.
Knot IDs are displayed in the H$\alpha$ intensity map.
(A color version of this figure is available in the online journal.)
\label{fig:image-3dii}}
\end{figure}

Figure \ref{fig:image-3dii} (a) displays the continuum surface brightness map
of the central part of the M82 cap.
Only stars in the Galaxy and distant galaxies are detected.
The number counts are consistent with the freely available GalaxyCount program
which provides source statistics for any window function down to 28 AB mag
\citep{Ellis:2007}.
The relative positions of objects in our image coincide with those of
\citet{Devine:1999} and SDSS DR7 \citep{Abazajian:2009}.
In contrast, we cannot detect continuum emission from the cap.
The upper limit in surface brightness at 6656 \AA\, is 23.7 mag arcsec$^{-2}$
(AB, 5$\sigma$), which corresponds to stellar mass of approximately 3 $\times$
10$^7$ $M_{\odot}$ using the cap size as 0.5 kpc$^2$ and mass-to-luminosity
ratio at solar metallicity, star-formation history of SSP, and an age of 1 Gyr
estimated from the \citet{Bruzual:2003} model.
This fact indicates that the cap is not a dwarf galaxy \citep{Lehnert:1999}.

In the H$\alpha$ intensity map (Figure \ref{fig:image-3dii} (b)), clumpy and
filamentary structures in the cap are clearly detected.
Our H$\alpha$ map is roughly consistent with that of \citet{Devine:1999}, but
bright knots C and D identified by them are not confirmed in our H$\alpha$
image; instead these are detected in our continuum image.
The H$\alpha$ image of \citet{Devine:1999} is clearly contaminated by continuum
emission.
Knots C and D are not related to the cap, and appear to be more distant disk
galaxies.
Our high resolution (= 0\arcsec.9) H$\alpha$ map enables us to resolve some
bright H$\alpha$ knots.
The typical size of H$\alpha$ knots is 5\arcsec--10\arcsec, which corresponds
to 90--180 pc at the distance of M82.

We renamed the H$\alpha$ knots after \citet{Devine:1999}, as shown in the
H$\alpha$ image (Figure \ref{fig:image-3dii} (b)).
The H$\alpha$ flux and luminosity of the brightest knot H-C in 5\arcsec\,
$\times$ 5\arcsec\, aperture are 1.7 $\times$ 10$^{-15}$ erg cm$^{-2}$ s$^{-1}$
and 2.8 $\times$ 10$^{36}$ erg s$^{-1}$, respectively.
The electron density and mass of the brightest knot H-C are estimated as 1.0 
$f_{\rm H\alpha,1}^{-1/2}$ cm$^{-3}$ and 9 $\times$ 10$^3
f_{\rm H\alpha,1}^{1/2} M_{\odot}$, respectively, where $f_{\rm H\alpha,1}$
indicates the filling factor of the knot.
They are estimated from its H$\alpha$ luminosity, size (5\arcsec\, = 90 pc),
and the H$\alpha$ recombination rate $\alpha_{\rm H\alpha}$ = 8.7 $\times$
10$^{-14}$ cm$^3$ s$^{-1}$ \citep{Osterbrock:2006}, with the assumption of
spherical symmetry and a completely ionized gas.
The observed H$\alpha$ flux at each knot is displayed in Table 
\ref{tb:data-knot}.
The total H$\alpha$ flux of the cap region in our field of view is 7.3 $\times$
10$^{-14}$ erg cm$^{-2}$ s$^{-1}$.
This flux is about half of that estimated by \citet{Devine:1999} and
\citet{Lehnert:1999} ($\sim$1.3 $\times$ 10$^{-13}$ erg cm$^{-2}$ s$^{-1}$)
consistent with the fact that about half of the cap region falls within our
field of view.

\begin{deluxetable}{cccc}
\tablecaption{Observed H$\alpha$ flux and line ratios at each knot
\label{tb:data-knot}}
\tablewidth{0pt}
\tablehead{
\colhead{Knot ID} & \colhead{H$\alpha$ flux} &
\colhead{[\ion{N}{2}]$\lambda$6583/H$\alpha$} &
\colhead{[\ion{S}{2}]$\lambda\lambda$6716,6731/H$\alpha$} \\
\colhead{} & \colhead{(10$^{-16}$ erg cm$^{-2}$ s$^{-1}$)} & \colhead{} &
\colhead{} 
}
\startdata
E-C & 4.3 $\pm$ 0.1 & 0.34 $\pm$ 0.03 & 0.63 $\pm$ 0.06 \\
E-N & 4.0 $\pm$ 0.1 & 0.22 $\pm$ 0.03 & 0.60 $\pm$ 0.06 \\
E-E1 & 3.3 $\pm$ 0.1 & 0.32 $\pm$ 0.04 & 0.63 $\pm$ 0.08 \\
E-E2 & 2.8 $\pm$ 0.1 & 0.27 $\pm$ 0.04 & 0.45 $\pm$ 0.08 \\
F-C & 3.2 $\pm$ 0.1 & 0.18 $\pm$ 0.03 & 0.55 $\pm$ 0.07 \\
F-N & 2.6 $\pm$ 0.1 & 0.20 $\pm$ 0.04 & 0.60 $\pm$ 0.09 \\
F-S & 2.5 $\pm$ 0.1 & 0.20 $\pm$ 0.05 & 0.55 $\pm$ 0.08 \\
G-C & 2.3 $\pm$ 0.1 & 0.21 $\pm$ 0.05 & 0.56 $\pm$ 0.10 \\
G-SE1 & 3.1 $\pm$ 0.1 & $<$ 0.11 & 0.63 $\pm$ 0.08 \\
G-SE2 & 3.1 $\pm$ 0.1 & 0.12 $\pm$ 0.04 & 0.62 $\pm$ 0.08 \\
G-SW & 2.9 $\pm$ 0.1 & 0.18 $\pm$ 0.04 & 0.52 $\pm$ 0.08 \\
H-C & 5.7 $\pm$ 0.1 & $<$ 0.06 & 0.55 $\pm$ 0.04 \\
H-W & 3.3 $\pm$ 0.1 & $<$ 0.10 & 0.66 $\pm$ 0.07 \\
J-C & 2.8 $\pm$ 0.1 & 0.24 $\pm$ 0.04 & 0.55 $\pm$ 0.09 \\
J-E & 2.9 $\pm$ 0.1 & $<$ 0.12 & 0.46 $\pm$ 0.07 \\
J-S & 2.6 $\pm$ 0.1 & 0.17 $\pm$ 0.04 & 0.50 $\pm$ 0.08 
\enddata
\tablecomments{The aperture sizes are 2\arcsec.25 $\times$ 2\arcsec.25.
Uncertainties are in 1 $\sigma$ levels, while upper limits are in 3 $\sigma$.}
\end{deluxetable}

\begin{figure}
\epsscale{1.00}
\plotone{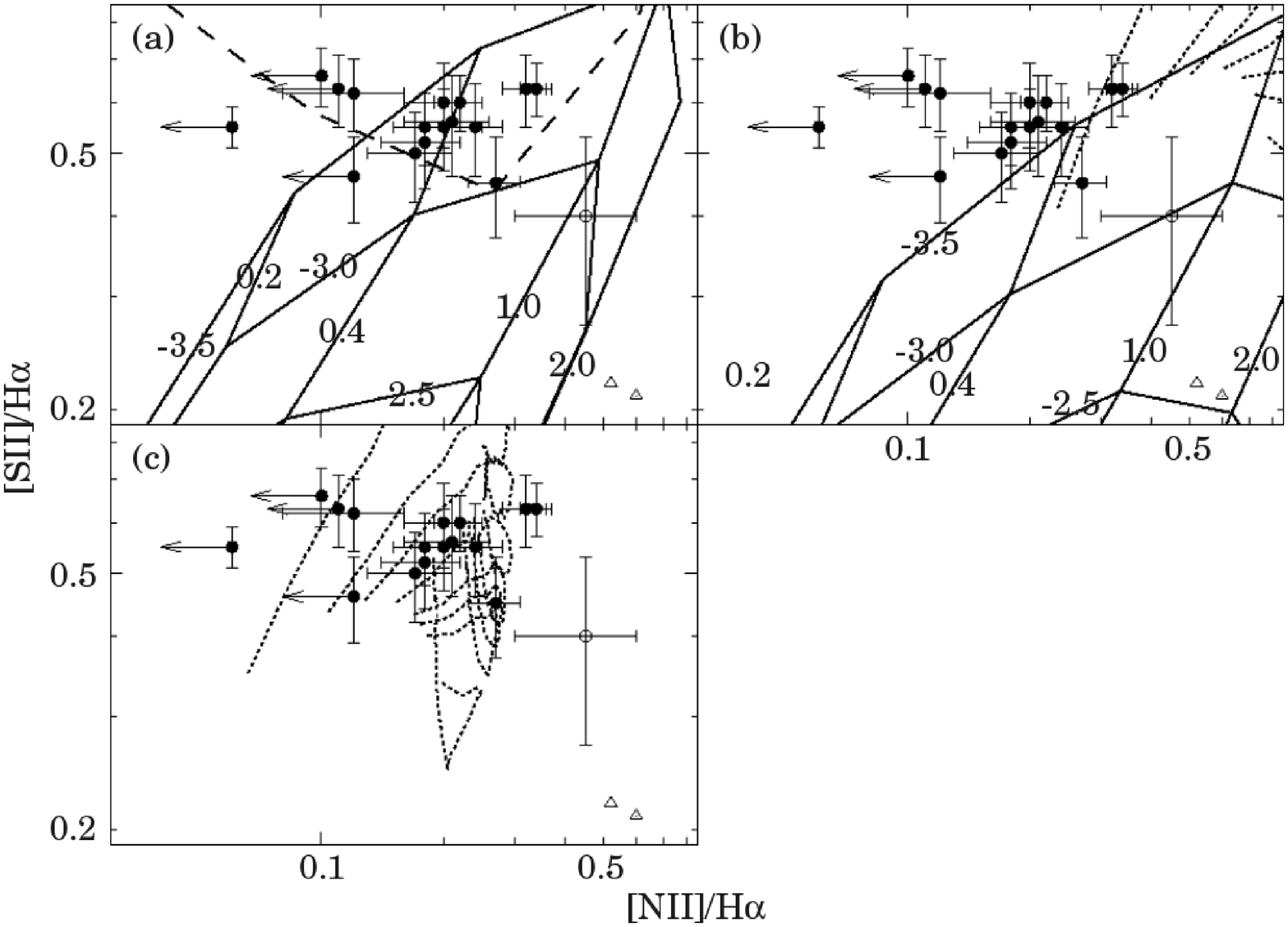}
\caption{Observed [\ion{S}{2}]/H$\alpha$ against [\ion{N}{2}]/H$\alpha$ ratios
listed in Table \ref{tb:data-knot}.
Filled circles, open circles, and open triangles represent the observed ratios
at M82 cap (this study), superwind regions $\sim$1 kpc from the nucleus
\citep{Shopbell:1998, Yoshida:2011}, and central regions \citep{O'Connell:1978,
Smith:2006}, respectively.
Uncertainties in this study are in 1 $\sigma$ levels, while upper limits are in
3 $\sigma$.
Errorbars on the superwind region data represent the ranges of line ratios in
their field of view.
Line ratio grids calculated by a photoionization model
\citep[{\it Cloudy}: ][]{Ferland:1998} are plotted in solid lines.
The assumed electron densities are (a) 1 cm$^{-3}$ and (b) 2000 cm$^{-3}$.
The positive and negative values on the model lines denote metallicity and
ionization parameter, respectively.
Dashed line in panel (a) displays the line ratios calculated by a slow shock
model of \citet{Shull:1979} with solar metallicity as a function of shock
velocity.
Data points for a shock velocity of 60, 80, and 90 km s$^{-1}$ appear in this
panel.
The dotted lines in panels (b) and (c) display the line ratios calculated by a
fast shock model of \citet{Allen:2008} with solar and LMC metallicities,
respectively, as a function of shock velocity at various magnetic field
strengths.
The left edge of each line corresponds to the lowest shock velocity, 200 km 
s$^{-1}$.
\label{fig:n2ha-s2ha}}
\end{figure}

Figures \ref{fig:image-3dii} (c), (d), (e), and (f) displays
[\ion{N}{2}]$\lambda$6583 intensity, [\ion{S}{2}]$\lambda\lambda$6716,6731
intensity, [\ion{N}{2}]/H$\alpha$ ratio, and [\ion{S}{2}]/H$\alpha$ ratio maps
of a part of the M82 cap, respectively.
The [\ion{S}{2}] flux map is generally similar to the H$\alpha$ map.
All counterparts of H$\alpha$ knots are also found in the [\ion{S}{2}] map.
[\ion{S}{2}]/H$\alpha$ flux ratios are almost constant among these knots,
0.45--0.66.
The [\ion{N}{2}] flux map is quite different from the H$\alpha$ flux map.
Knots E-C and E-E are clearly detected in [\ion{N}{2}], but the counterparts of
the other knots, even the H$\alpha$ brightest knot H-C, are barely detected.
The observed [\ion{N}{2}]/H$\alpha$ flux ratios of knots E-C and E-E1 are the
largest and peak at $\sim$0.33.
[\ion{N}{2}]/H$\alpha$ and [\ion{S}{2}]/H$\alpha$ line ratios at these knots
are summarized in Table \ref{tb:data-knot} and plotted in Figure
\ref{fig:n2ha-s2ha}.
Small [\ion{N}{2}]/H$\alpha$ and [\ion{S}{2}]/H$\alpha$ ratios are consistent
with the previous result of non-detection of forbidden lines at the cap
\citep{Devine:1999}.
We do not find correlations between the H$\alpha$ flux, [\ion{N}{2}]/H$\alpha$
and [\ion{S}{2}]/H$\alpha$ ratios.

We compare the [\ion{N}{2}]/H$\alpha$ and [\ion{S}{2}]/H$\alpha$ ratios of the 
cap with those at other regions in M82.
The emission line ratios at a radius of 1 kpc from M82's center
\citep{Shopbell:1998, Yoshida:2011}, and in the circumnuclear regions
\citep{O'Connell:1978, Smith:2006}, are also plotted in Figure
\ref{fig:n2ha-s2ha}.
We find an interesting trend of the line ratios with galactic radius.
The [\ion{N}{2}]/H$\alpha$ ratio tends to decrease, while the
[\ion{S}{2}]/H$\alpha$ ratio tends to increase with distance from the M82
nucleus.
This fact suggests that some parameters, such as metallicity and shock
velocity, gradually change, if the ionization source is the same in these
regions.
In Figure \ref{fig:n2ha-s2ha}, the line ratios of the M82 starburst regions and
the cap are significantly different.
Therefore, dust reflection of the M82 starburst regions, suggested by strong UV
emission \citep{Hoopes:2005}, is not the dominant emission mechanism in optical
wavelength.

\begin{figure}
\epsscale{1.00}
\plotone{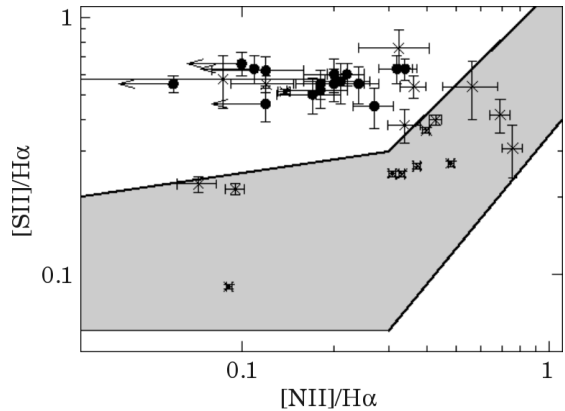}
\caption{Observed [\ion{N}{2}]/H$\alpha$ vs [\ion{S}{2}]/H$\alpha$ of M82 cap
(filled circle) and EIGs in the Coma cluster (cross) \citep{Yoshida:2012}.
The shaded region represents the line ratios of \ion{H}{2} regions of nearby
field galaxies \citep{Jansen:2000}, blue compact galaxies \citep{Kong:2002},
and LINERs \citep{Ho:1997}.
\label{fig:n2ha-s2ha-coma}}
\end{figure}

Figure \ref{fig:n2ha-s2ha-coma} compares the observed [\ion{N}{2}]/H$\alpha$
and [\ion{S}{2}]/H$\alpha$ ratios of the M82 cap with those of \ion{H}{2}
regions \citep{Jansen:2000}, blue compact galaxies \citep{Kong:2002}, LINERs
\citep{Ho:1997}, and very extended ionized gas (EIG) in the Coma cluster
\citep{Yoshida:2012}.
We find that the line ratios of the M82 cap are similar to those of some EIG
knots which have larger [\ion{S}{2}]/H$\alpha$ ratios than the main sequence of
star-forming galaxies.
This fact suggests that the emission line ratios of the M82 cap are not
peculiar, and that the M82 cap and EIGs in the Coma cluster are ionized by the
same mechanism.

\section{DISCUSSION}
\label{sec:discuss}

First we discuss the dominant ionization source of the M82 cap.
The observed [\ion{N}{2}]/H$\alpha$ and [\ion{S}{2}]/H$\alpha$ line ratios 
fall within the range of star-forming galaxies (e.g., \citealt{Liang:2006}) and
diffuse ionized gas (e.g., \citealt{Moustakas:2006}).
The ionization source of star-forming galaxies is UV photons from massive stars
(e.g., \citealt{Kewley:2001}), whereas there are some possible sources for
diffuse ionized gas and EIG, such as photoionization, shock (e.g.,
\citealt{Allen:2008}), and turbulent mixing layers (e.g.,
\citealt{Slavin:1993}).
In the case of turbulent mixing layers, the H$\alpha$ emitting ionized gas
should exist at the boundary of the hot gas, but they appear to be spatially
coincident; thus, we rule out this model.
In order to reveal the ionization source of the cap, we compare the observed
line ratios with theoretical values for photoionization and shock models.

\subsection{Photoionization}

Photoionization of the cap by the M82 nuclear starburst region was suggested 
by \citet{Lehnert:1999}.
They calculated the number of ionizing photons at the cap region, and found
that there are enough ionizing photons relative to that estimated from
H$\alpha$ luminosity.
This model assumes that ionizing photons are not strongly absorbed or scattered
by interstellar matter between the M82 starburst regions and the cap region,
presumably because the wind phenomenon has cleared the sight line of obscuring
matter.

The observed [\ion{N}{2}]/H$\alpha$ and [\ion{S}{2}]/H$\alpha$ ratios (Table 
\ref{tb:data-knot}) are similar to those of \ion{H}{2} galaxies (e.g.
\citealt{Liang:2006}), and the observed ratios do not exceed the maximum
starburst line defined by \citet{Kewley:2001}, unless the
[\ion{O}{3}]$\lambda$5007/H$\beta$ ratio is larger than $\sim$ 3.
But the observed [\ion{S}{2}]/H$\alpha$ ratio is roughly equal to
the largest value of $\sim$ 40,000 SDSS star-forming galaxies
\citep{Liang:2006} making UV photoionization less probable.

In order to clarify whether photoionization can produce the observed larger
[\ion{S}{2}]/H$\alpha$ ratios, we calculated emission line ratios with a
{\it Cloudy} photoionization model \citep{Ferland:1998}.
We used two electron densities for the calculation: 1 cm$^{-3}$ for the cap,
and 2000 cm$^{-3}$ for the M82 center \citep{O'Connell:1978, Smith:2006}.
Due to its secondary nature, we assume that the nitrogen abundance scales with
metallicity ($Z_{\rm gas}$) as $Z^2_{\rm gas}$ and that the stellar metallicity
($Z_\star$) is the same as $Z_{\rm gas}$ (see \citealt{Nagao:2011}).
The SED of the stars is taken from \citet{Leitherer:1999}.

Figure \ref{fig:n2ha-s2ha} (a) indicates that most of the observed line ratios
at the M82 cap are reproduced by the photoionization model of ionization
parameter $U$ = 10$^{-3.5}$--10$^{-3.0}$ and metallicity $Z \sim$ 0.4
$Z_{\odot}$.
$U$ represents the dimensionless ratio of the ionizing photon density to the
electron density.
To constrain the ionization parameter $U$ of the M82 cap, we follow the
\citet{Lehnert:1999} condition: the dimension of 3.7 $\times$ 3.7 kpc$^2$,
electron density of 4.3 $\times$ 10$^{-2} f_{\rm H\alpha,2}^{-1/2}$ cm$^{-3}$,
where $f_{\rm H\alpha,2}$ indicates the filling factor of ionized gas averaged
over the whole M82 cap, and the number of ionizing photons at the cap of 2
$\times$ 10$^{50}$ photon s$^{-1}$.
The calculated ionization parameter $U = 1.2 \times 10^{-3} 
f_{\rm H\alpha,2}^{1/2}$ is consistent with that estimated from emission line
ratios, if we assume $f_{\rm H\alpha,2} = 1$.
The larger distance from the ionization source than for typical \ion{H}{2}
regions leads to smaller ionization parameter and larger [\ion{S}{2}]/H$\alpha$
ratios.

The number of the ionizing photons which each knot receives is almost the same,
and the ionization parameter depends on the knot electron density only.
The calculated line ratios whose electron densities are smaller than 10
cm$^{-3}$ are almost the same.
For these reasons, we can consider that the difference in line ratios among the
knots in the cap is explained by the difference in the electron density and/or
metallicity, as seen in Figure \ref{fig:n2ha-s2ha} (a).
Although the lower density model cannot reproduce the ratios at the central
regions, log $U$ = $-2.5$ and $Z$ = 2.0 of the higher density model fits the
observed ratios (Figure \ref{fig:n2ha-s2ha} (b)).

However, taking account of the escape fraction of ionizing photons, we find
that this picture cannot be correct.
\citet{Lehnert:1999} uses H$\alpha$ surface brightness averaged over the
projected length of the M82 cap for the calculation of the escape fraction of
ionizing photons.
Since the required escape fraction is only $\sim$3 \%, they considered that
the M82 starburst regions can provide enough ionizing photons for the M82 cap.
Meanwhile, our high resolution image reveals that M82 cap is patchy and its
filling factor $f_{\rm H\alpha,2}$ is much smaller than 1 (Figure
\ref{fig:image-3dii} (b)).
This means that the H$\alpha$ surface brightness of the brightest knot H-C must
be explained by the ionizing photon flux from the M82 starburst region.
Knot H-C is smaller and denser than the whole M82 cap, and therefore much
larger escape fraction is required.
We estimate the required escape fraction of ionizing photons from the M82
starburst regions.
We assume that the knots are spherical symmetric (Figure \ref{fig:image-3dii}).
Then we can use the same calculation method as \citet{Bland-Hawthorn:1999,
Bland-Hawthorn:2002}.
The H$\alpha$ surface brightness of knot H-C is 1 $\times$ 10$^{-16}$ erg
cm$^{-2}$ s$^{-1}$ arcsec$^{-2}$ in 2\arcsec.25 $\times$ 2\arcsec.25 aperture
(Table \ref{tb:data-knot}).

Since the surface brightness of 6 $\times$ 10$^{-18}$ erg cm$^{-2}$ s$^{-1}$
arcsec$^{-2}$ corresponds to 1 Rayleigh at the H$\alpha$ wavelength, the
H$\alpha$ emission measure of the knot H-C is 16.5 Rayleigh.
The required number of ionizing photons is 3.7 $\times$ 10$^7$ cm$^{-2}$
s$^{-1}$ at the distance of M82, using the H$\alpha$ recombination rate
$\alpha_{\rm H\alpha}$ = 8.7 $\times$ 10$^{-14}$ cm$^3$ s$^{-1}$
\citep{Osterbrock:2006}.
Whereas the ionizing photon flux from the nuclear starburst region is 10$^{54}$
photons s$^{-1}$ \citep{McLeod:1993}, and 6 $\times$ 10$^7$ cm$^{-2}$ s$^{-1}$
at the distance of knot H-C from the M82 nuclear starburst region. Thus
the required escape fraction for knot H-C is 60 \%.
This escape fraction is an order of magnitude larger than what has been 
measured in the Galaxy ($\sim$6 \%: \citealt{Bland-Hawthorn:1999}) and dwarfs
to date ($\sim$3 \%: \citealt{Zastrow:2011, Barger:2012}).
In passing, we note that 3D simulations of UV radiative transfer in superwinds
indicate that higher values are possible in extreme cases \citep{Yajima:2009}. 
Thus, it is quite unlikely that the M82 cap clouds are ionized by photons from
the M82 starburst regions.
We consider shock ionization to be a more likely explanation, as we discuss in
the next section.

\subsection{Shock ionization}

In order to explain the H$\alpha$ emission in the cap, \citet{Lehnert:1999}
first suggested that the M82 superwind can drive a shock into the underlying
gas cloud. 
The gas metallicity, which is presently unknown, is likely to fall in the range
0.1--1 $Z_\odot$ depending on whether the gas is infalling or entrained with
the wind flow.
The threshold metallicity for infalling gas at the present epoch appears to be
close to 0.1$Z_\odot$ in all observations of the Local Universe to date
\citep{Ryan-Weber:2009}.
We regard this value as a lower limit because the outer \ion{H}{1} envelope in
the M81 group appears to be material stripped from the outer disk of one or
more galaxies \citep{Chynoweth:2007}.
The cap is unlikely to be entrained gas from the disk because dense
material is broken up very quickly by Rayleigh-Taylor instabilities
\citep{Cooper:2009}, as observed in the wind filaments close to the disk.

We compare the observed line ratios with those calculated from a fast shock
model ($v_{\rm s,1} \geq$ 200 km s$^{-1}$) of \citet{Allen:2008}.
The [\ion{N}{2}]/H$\alpha$ ratios in the shock model are larger than 0.3, while
the observed values are mostly less than 0.3 (Figure \ref{fig:n2ha-s2ha} (b)).
But this may reflect the lower expected metallicity in the cap.
Fast shocks in a low metallicity gas may be able to explain the observed
[\ion{N}{2}]/H$\alpha$ ratios.
The computed [\ion{N}{2}]/H$\alpha$ ratios fall in the range 0.1 to 0.3, which
is similar to the observed values.
But if fast shock excitation is dominant, the observed [\ion{N}{2}]/H$\alpha$
ratios should correlate with [\ion{S}{2}]/H$\alpha$, because both ratios
increase in lock step with an increase in shock velocity (Figure
\ref{fig:n2ha-s2ha} (c)); however, no such correlation between these ratios is 
found.
A more compelling argument against fast shocks is the kinematically `cold' line
emission observed across the `cap' region, an issue we return to below. 
Thus fast shocks are unlikely to be the dominant ionization source of the cap
today.

Next we compare the observed ratios with those calculated from slow to
intermediate shock models, i.e., 40 km s$^{-1} \leq v_{\rm s,1} \leq$ 130 km
s$^{-1}$, given by \citet{Shull:1979} (Figure \ref{fig:n2ha-s2ha} (a)).
In this shock velocity range, [\ion{N}{2}]/H$\alpha$ ratios increase as shock
velocity increases, while [\ion{S}{2}]/H$\alpha$ ratios neither increase nor
decrease monotonically.
In Figure \ref{fig:n2ha-s2ha} (a), the observed points at the cap knots are
distributed along the line of slow shock model.
The calculated [\ion{N}{2}]/H$\alpha$ and [\ion{S}{2}]/H$\alpha$ ratios are
0.10 and 0.58 at $v_{\rm s,1}$ = 60 km s$^{-1}$, and 0.27 and 0.44 at
$v_{\rm s,1}$ = 80 km s$^{-1}$, respectively. 
Therefore, shock velocities of 60 km s$^{-1}$ and 80 km s$^{-1}$ can also
explain the observed [\ion{N}{2}]/H$\alpha$ and [\ion{S}{2}]/H$\alpha$ ratios.
Model comparisons for shock velocities higher than 80 km s$^{-1}$ are ruled
out.

\begin{figure}
\epsscale{1.00}
\plotone{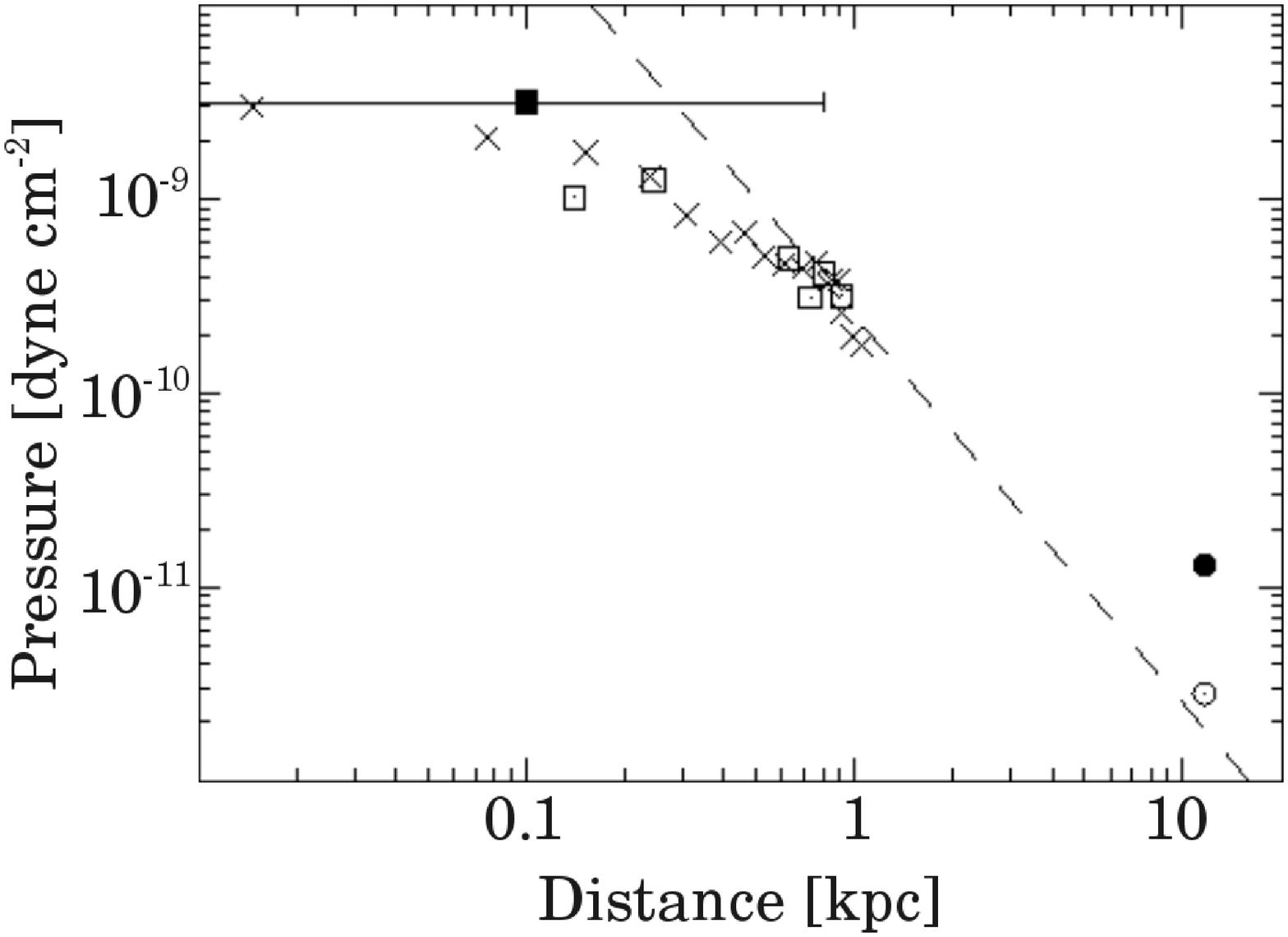}
\caption{Radial pressure profile along the M82 minor axis.
Crosses, open squares, and an open circle represent the thermal pressure of
ionized gas from \citet{Heckman:1990}, \citet{Yoshida:2011}, and this study,
respectively.
The temperature of ionized gas is assumed to be 10$^4$ K.
A filled square and a filled circle represent the thermal pressure of hot gas
from \citet{Strickland:2007} and \citet{Lehnert:1999}, respectively.
The filling factor of hot gas is assumed to be unity.
Thermal pressures are calculated by $2nk_{\rm B}T$ in both cases, where $n$ and
$k_{\rm B}$ are the electron density of ionized gas and Boltzmann's constant,
respectively.
The slope of the dashed line is given by $r^{-2}$, where $r$ is the distance
from the M82 center.
\label{fig:pressure-profile}}
\end{figure}

The superwind is powered by kinetic energy from some combination of stellar
winds, radiation pressure and supernovae in the M82 starburst region.
Here we examine whether the momentum is enough to produce the observed shock.
We compare the inferred range of shock velocities (40--80 km s$^{-1}$) with
the superwind velocity of a spherically-symmetric model for M82 by
\citet{Chevalier:1985}.
Their model fits well with the observed thermal pressure profile of M82 within
$\sim$1 kpc, and the pressure profiles at the larger radii than 0.5 kpc are
consistent with the $r^{-2}$ dependence expected for ram pressure
\citep{Heckman:1990}.
Using their model, the gas density, wind velocity, and ram pressure at the
distance of M82 cap are estimated as 4 $\times$ 10$^{-6}$ cm$^{-3}$, 5600 km
s$^{-1}$, and 2.1 $\times$ 10$^{-12}$ dyne cm$^{-2}$, respectively.
Since the observed electron density is 1.0 cm$^{-3}$ at the knot H-C, the
thermal pressure of ionized gas is calculated as $2nk_{\rm B}T = 2.8 \times
10^{-12}$ dyne cm$^{-2}$ with the assumption of $T = 10^4$ K.
This indicates that the model ram pressure and the observed thermal pressure
are well balanced, and the observed pressure follows $r^{-2}$ law even at the
distance of the M82 cap, 11.6 kpc (Figure \ref{fig:pressure-profile}).
Furthermore, given that some of the shock motion is expected to generate local
turbulent motions of the same order \citep{Bland-Hawthorn:2007}, the measured
narrow emission line widths ($\lesssim 100$ km s$^{-1}$ FWHM,
\citealt{Devine:1999}) are consistent with a slow shock.
Therefore, it is quite likely that the cap clouds are ionized by a slow shock
produced by the M82 superwind.

Future deep observations of the M82 cap, e.g., [\ion{O}{1}]$\lambda$6300, will
enable us to directly confirm its dominant ionization source.
In the case of a slow shock, the [\ion{O}{1}]/H$\alpha$ ratio larger than 0.1
is to be expected \citep{Shull:1979}, while a smaller ratio is evidence for
photoionization by massive stars (e.g., \citealt{Kewley:2001}).

\subsection{Hot gas heating}

Diffuse X-ray emission was detected at the M82 cap \citep{Devine:1999,
Lehnert:1999, Tsuru:2007}.
The electron density, temperature, and thermal pressure of hot gas are $5.4
\times 10^{-3} f_{\rm X}^{-1/2}$ cm$^{-3}$, 9 $\times 10^6$ K, and 1.3 $\times
10^{-11} f_{\rm X}^{-1/2}$ dyne cm$^{-2}$, respectively, where $f_{\rm X}$ is
the filling factor of hot gas.
However, a slow shock is not the main heating mechanism, because these can only
heat to $\sim10^5$ K.
Additionally, thermal pressures of ionized and hot gas are inconsistent at the
M82 cap, though they are consistent at the M82 center \citep{Heckman:1990,
Strickland:2007}.
Hence, another heating source for hot gas is required.

A reasonable explanation for the hot gas is found when considering a
2-component shock model.
An \ion{H}{1} cloud in isolation develops a core-halo structure where the core
is dense and the halo is relatively diffuse \citep{Field:1965, Sternberg:2002}.
In this interpretation, the X-ray emission is produced by the superwind
triggering a shock in the diffuse gas surrounding the knots in the cap (see
\citealt{Lehnert:1999}).
To produce gas with $T = 9 \times 10^6$ K requires a fast shock whose shock
velocity $v_{s,2}$ is 820 km s$^{-1}$.
But the 2-component shock by the same superwind cannot explain the discrepancy
between the thermal pressures of these gas phases.

A 2-component shock by different superwinds from the M82 starburst region may
be the case for the M82 cap.
In this model, the slow shock by the present superwind ionizes the cap clouds,
while hot gas observed today was produced by a fast shock driven by a past
superwind outburst.
What is important is that the cooling timescales of these gas phases are
drastically different: $\approx10^4$ yr and $\approx10^8$ yr for ionized and
hot gas, respectively \citep{Osterbrock:2006, Lehnert:1999}.
The history of this model is as follows: $\lesssim10^8$ yrs ago, a fast shock
driven by a past superwind produced hot ionized gas; after $\approx10^4$ yrs,
the ionized gas cooled while the hot gas continued to emit X-rays, and now the
slow shock ionizes the cap clouds again to produce the H$\alpha$ and X-ray
emission from the cap.

\section{Concluding Remarks}

It is undoubtedly true that the physics of superwinds is complicated.
However, we are able to deduce some basic properties of the wind.
As discussed by \citet{Sharp:2010}, there is a well determined series of events
that lead to a superwind taking hold.
We believe that the new observations support the developing paradigm.

When a critical surface density of dense molecular clouds is reached in the
galaxy, massive stars are born.
These stars evolve rapidly and their strong UV radiation fields produce a warm
gaseous medium that encircles the remaining molecular clouds.
After a few million years, the cores of the most massive stars collapse leading
to multiple supernova explosions.
This huge impulse of mechanical energy heats the diffuse medium to extremely
high temperatures ($T\sim 10^8$K) where it expands to form a powerful
superwind.
(If the mechanical energy or UV radiative energy were entirely absorbed by the
dense gas, most of the energy would simply be re-radiated as IR emission.)
The hot flowing gas entrains cooler gas from the disk with the flow; we know
this because of well defined rotation of the entrained filaments about the wind
axis \citep{Greve:2004}.
The entrained gas is very clumpy because it is mostly entrained from the
surviving dense clouds in the disk \citep{Cooper:2008}. 

M82 is engulfed by a large \ion{H}{1} cloud complex and some of this material
appears to be accreting onto the dwarf close to the minor axis 
\citep{Chynoweth:2007}.
The cap material is almost certainly supplied by infalling gas.
The cap ionization almost certainly arises from the expanding superwind
interacting with infalling gas.
The spatial coincidence of the H$\alpha$ and X-ray emission may be explained
with the model of a 2-component shock by different M82 superwinds.
Further constraints on the cap may come from UV absorption line spectroscopy
using distant quasars as a background light source.
The expected warm-hot medium should be visible in low to intermediate
ionization states of C, N, O, Ne and Fe, even for gas in a non-equilibrium
state \citep{Gnat:2004}.
Our new highly resolved observations show that the cap is very clumpy
presumably as a consequence of the wind-cloud interaction.
Our expectation is that the cap will be disrupted in a shock crossing time of
about $\sim 1$ Myr.

The fact that the outer cap is presently ionized by local shocks, and not by
nuclear UV radiation, supports the findings of \citet{Sharp:2010}.
Unlike AGN-driven wind filaments which are all found to be ionized by nuclear
UV radiation, the starburst wind filaments are ionized by local shocks far
from the nucleus.
This is easily understood in terms of a `starburst' because the hot young stars
must evolve to supernovae before the wind gets going, and therefore few remain
to ionize gas clouds in the direction of the flow.
We fully anticipate that future deep optical, UV and X-ray imaging and
spectroscopy will reveal further details about the cap region, and in turn
about the nature of the superwind in M82.

\acknowledgments

This work is based on data collected at Subaru Telescope, which is operated by
the National Astronomical Observatory of Japan.
We thank M. Yoshida for useful discussions.
This work was financially supported in part by the Japan Society for the
Promotion of Science (Nos. 17253001, 19340046, 23244031, and 23654068).

\end{document}